\documentclass{ckm}                 
\usepackage{txfonts}            

\confname{Workshop on the CKM Unitarity 
Triangle, IPPP 
Durham, April
  2003; hep-ph/0306058}
\newcommand{\lsim}{
\mathrel{\hbox{\rlap{\hbox{\lower4pt\hbox{$\sim$}}}\hbox{$<$}}}}

\title{Determination of $\gamma$ and $\alpha$ from non-leptonic
B decays\\ with SU(3) flavour symmetry}

\author{Joaquim Matias}
\address{IFAE, Universitat Aut\`onoma de Barcelona, 08193
Bellaterra, Barcelona, Spain}

\begin{document}

\begin{abstract}We describe in detail the method we have used  to determine the
CKM angles $\gamma$, $\alpha$ and $\beta$ using flavour
symmetries between non-leptonic B decays. This method is valid 
in the context of the SM but
also in presence  of New Physics not affecting the amplitudes.
\end{abstract}

\maketitle


\section{Introduction}

B factories are opening a new exciting period in the precision
Flavour Physics \cite{first}. A set of very interesting
non-leptonic B
decays: $B \to \pi K$ and $B\to \pi \pi$ are
now accessible at the $e^+e^-$ B factories. These modes together
with the CP-asymmetry of $B_d \to J/\Psi K_S$ will allow us to
determine the CKM angles, $\gamma$, $\alpha$ and $\beta$.

In this talk we  try to answer the question of how precise we
can get to determine
the CKM angles using as inputs experimental data and symmetries, 
and trying
to minimize as much as possible hadronic uncertainties
from QCD. Since data seems to indicate that penguin diagrams play a
fundamental role, any method should include their 
contribution\cite{FlMa1,FlMa2,ot}.  We shall discuss, here, the
method we have used in\cite{FlMa1,FlMa2,FIM}
to determine the CKM angles. This method is based on
flavour symmetries between non-leptonic B decays.
Another very interesting approach to non-leptonic B decays in
the literature tries
to predict directly from QCD some of the hadronic parameters,
like, for instance, QCD
Factorization\cite{beneke} and PQCD\cite{sanda}.

We  focus, here, on the recently measured CP-violating 
 $B_d \to \pi \pi$ observables. We  construct the 
method, step 
by step, with emphasis on
its advantages and how to
improve it when data from hadronic machines\cite{hadronic} will
be available.
We follow the notation of\cite{FlMa2,FIM}.

\section{Description of the Method}

We  start by writing down the most general parametrization in the
SM of the amplitude corresponding to $B_d^0 \to \pi^+ \pi^-$, 
using the Wolfenstein
parametrization\cite{fl1,FlMa2}:

 $\begin{array}{ll}
{A(B_d^0 \rightarrow \pi^+
\pi^-)}&=
\lambda_u^{(d)}\left({ A_{\rm
CC}^{u}}+{A_{\rm pen}^{(u)}} \right)
+ \lambda_c^{(d)} { A_{\rm
pen}^{(c)}}+\lambda_t^{(d)} { A_{\rm
pen}^{(t)}}
\nonumber \cr \nonumber \cr
&={\cal
C}\left(e^{i
\gamma} -  {\bf d
e^{i\theta}}\right) \nonumber
\end{array}$

This amplitude includes current-current contributions and QCD 
and EW penguin diagrams.
All the hadronic information is collected in:
\begin{equation}\label{dt} {\bf d
e^{i\theta}}\equiv\frac{1}{R_b}\left(
\frac{{ A_{\rm pen}^{ct}}}{{
A_{\rm
CC}^{u}}+{ A_{\rm pen}^{ut}}}\right),  \quad
{\cal C}\equiv\lambda^3A\,R_b\left({ A_{\rm
CC}^{u}}+{ A_{\rm
pen}^{ut}}\right)\end{equation}
with
$A\equiv|V_{cb}|/\lambda^2$, 
$R_b=(1-\lambda^2/2)|V_{ub}/\lambda V_{cb}|$ and $A_{\rm 
pen}^{qt}\equiv A_{\rm pen}^{q}-A_{\rm pen}^{t}$.
We can construct, using this amplitude, the  
direct and mixing induced
CP-asymmetries of $B_d \to \pi^+ \pi^-$:
$$
{\cal A}_{\rm
CP}^{{\rm dir}}= -\left[\frac{2 {\bf
d}\sin{\bf\theta}\sin{\gamma}}{1- 2 {\bf
d}\cos{\bf\theta}\cos{\gamma }+{\bf
d^2}}\right]\label{ACP-dir}\nonumber
$$
$$
{{\cal A}_{\rm
CP}^{{\rm mix}}}=
\frac{\sin({\phi_d}+2{\gamma})-2
{\bf d} \cos\theta
\sin({\phi_d}+{\gamma})+ {\bf
d^2}\sin{\phi_d}}{1-2 {\bf d}
\cos\theta\cos{\gamma}+{\bf d^2}}\label{ACP-mix}
\nonumber
$$
Here, the  counting of parameters shows that we have two hadronic
parameters $\bf d$ and
$\theta$ and two weak parameters: weak mixing phase $\phi_d$ and
$\gamma$, but
only two observables.

However, we know that there is a closely related process $B_s \to KK$, where
a 
similar description can be used. A general amplitude 
parametrization\cite{fl1,FlMa2} in the SM is:
$$
{A(B_s^0
\rightarrow K^+ K^-)=}
\left(\frac{\lambda}{1-\lambda^2/2}\right) {\cal
C}{ '}\left[e^{i
\gamma}+\left(\frac{1-\lambda^2}{\lambda^2}\right)
{\bf d^\prime} e^{i\theta'}\right]
$$
It contains  also two hadronic parameters $\bf d^\prime$ and
$\theta^\prime$, with the
same functional dependence of penguin diagrams as in Eq.(\ref{dt}),
with the
only difference that the quarks $d$ and $s$ are interchanged in
the
external legs of penguins.

The corresponding CP asymmetries of $B_s \to K^+K^-$ are:
$$
{ {\cal A}_{\rm CP}^{{\rm dir}}}= \frac{2\tilde {\bf
d'}\sin\theta'\sin{\gamma}}{1+ 2\tilde {\bf
d'}\cos\theta'\cos{\gamma}+\tilde
 {\bf d'^2}}
$$
$$
 {{\cal A}_{\rm CP}^{{\rm mix}}}=
\frac{\sin({\phi_s}+2{\gamma})+2\tilde {\bf
d'}\cos\theta'\sin({\phi_s}+{\gamma})+ \tilde {\bf
d'^2}\sin{\phi_s}}{1+2\tilde {\bf d'}\cos\theta'\cos{\gamma}+
\tilde {\bf d'^2}}\label{ACPs-mix}
$$
They also depend  on two hadronic parameters: $\tilde {\bf 
d^\prime}
\equiv {\bf d^\prime}/\epsilon$ with $\epsilon \equiv {\lambda^2 /
(1 - \lambda^2)} \sim 0.05$ and $\theta^\prime$ and two weak
parameters: $\phi_s$ (negligibly small in SM) and the CKM angle 
$\gamma$.

Finally, if we combine both processes and their parameters using
the U-spin symmetry\cite{fl1,FlMa1,FlMa2}, that implies:
$${\bf d} e^{i \theta}={\bf d^\prime} e^{i \theta^\prime}$$
we will have four observables and five parameters (out of the initial seven):
 $\gamma$, $\phi_d$, 
$\bf d$, $\theta$ and $\phi_s$. Moreover, $\phi_s$  will be
determined from  ${\cal A}_{\rm CP}(B_s\to J/\Psi \phi)$.

Last but not least, we can  test the  U spin symmetry breaking in two
different ways:

a) One can define U-spin breaking parameters:
$\xi=\bf d^\prime/d$
and
$\Delta \theta=\theta^\prime-\theta$ and test the
sensitivity of the results to these parameters.

b) Once the data from $B_s \to KK$ will be available and 
${\cal A}_{\rm CP}(B_s\to J/\Psi \phi)$ measured ($\phi_s$), 
we will be able to reduce 
to three the number of parameters: $\gamma$, $\bf d$ and 
$\theta$, since ($\phi_d$ 
is taken from $B_d
\to J/\Psi K_S)$, so we can test $\xi$ or $\Delta \theta$.

Looking a bit more in detail one  realizes that  $\bf d$
is not a fully free parameter, we can constrain, and indeed
substitute it introducing a new observable called H (see
\cite{RF-Bpipi,FlMa2}):
$$
H\propto
\left[\frac{\mbox{BR}(B_d\to\pi^+\pi^-)}{\mbox{BR}(B_s\to
K^+K^-)}\right]
$$
that in the U-spin limit depends only on $\cos\theta\cos{\gamma}$
and $\bf d$. Although data on $B_s \to KK$ is still not
available, we can already now
apply the method using the data from B-factories, using the 
observation that $B_d \to \pi^\pm K^\mp$ and $ B_s \to K^+ K^-$
differ  in their  spectator quarks, meaning:
\begin{eqnarray} \label{spec}\phantom{center}
{\cal A}_{\rm CP}^{\rm dir}(B_s\to K^+K^-)&\approx&{\cal A}_{\rm
CP}^{\rm dir}
(B_d\to\pi^\mp K^\pm) \nonumber \cr
\mbox{BR}(B_s\to K^+K^-)
&\approx&\mbox{BR}(B_d\to\pi^\mp
K^\pm)\,\frac{\tau_{B_s}}{\tau_{B_d}}
\end{eqnarray}
This relation requires that the "exchange" and "penguin
annihilation" contributions to $B_s \to KK$ absent in $B_d \to
\pi^\pm K^\mp$ play a minor role\cite{ghdr}. 
But in case they would be enhanced we can also
control them  through data on 
$B_s
\to \pi^+\pi^-$.
 This allows us to determine  now H yielding:
$$ H\approx\frac{1}{\epsilon}\left(\frac{f_K}{f_\pi}\right)^2
\left[\frac{\mbox{BR}(B_d\to\pi^+\pi^-)}{\mbox{BR}(B_d\to\pi^\mp
K^\pm)} \right]=7.5 \pm 0.9
$$
and use it  to write $\bf d$ in terms of  $ {\bf
d}=f(
H,\theta,\gamma;\; \xi,\Delta \theta)$\cite{FlMa2}.

\section{Exploring the allowed region in $B_d\to\pi\pi$ to the
SM and beyond}
\begin{figure}[t]
\hbox to\hsize{\hss
\includegraphics[width=\hsize]{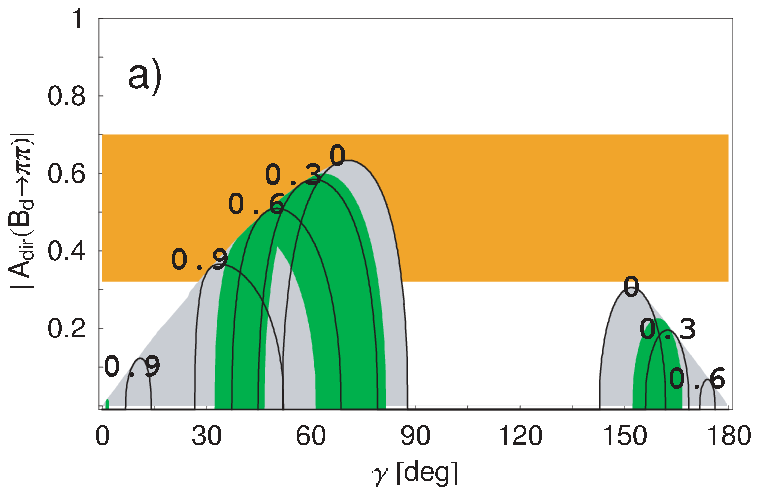}
\hss}
\includegraphics[width=\hsize]{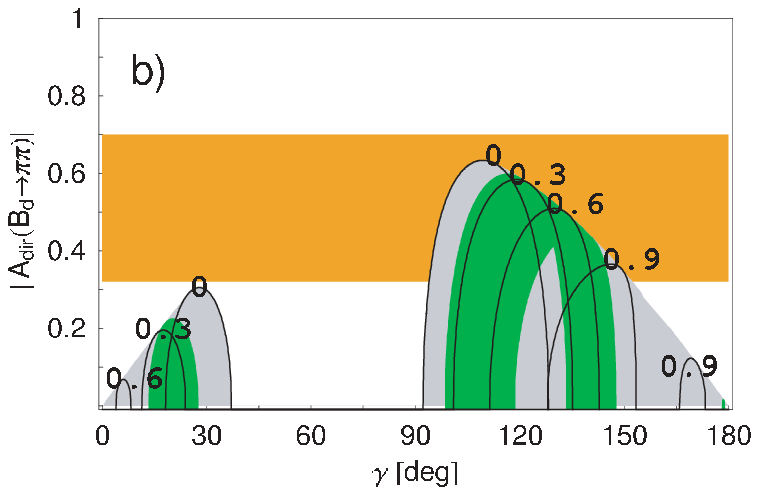}
\caption{$A_{CP}^{\rm dir}(B_d \to \pi\pi)$ as a function of 
$\gamma$   for
$A_{CP}^{\rm mix}(B_d \to \pi\pi)\in [0,1]$ and $H=7.5$. The 
curves correspond to
 fixed values for $A_{CP}^{\rm mix}(B_d \to \pi\pi)$. a) 
corresponds to the
solution $\phi_d=47^\circ$ and b) $\phi_d=133^\circ$. Horizontal
band correspond to the experimental value for $A_{CP}^{\rm dir}$
while the internal grey-shaded region 
corresponds to the
experimental value for $A_{CP}^{\rm mix}(B_d \to \pi\pi).$
 } \label{fig:cathedral}
\end{figure}
The starting point is the general expression\cite{FlMa2}:
\begin{equation}{ {\cal A}_{\rm CP}^{\rm
dir}(B_d\to\pi^+\pi^-)=}\mp\left[
\frac{\sqrt{4p^2-\left(u+vp^2\right)^2}\sin{\gamma}}{(1-
u\cos{\gamma})+ (1-v\cos{\gamma})p^2}\right]
\label{sp}
\end{equation}
where $u,v,p$ are functions of four observable quantities  ${\cal
A}_{\rm CP}^{\rm mix}$, $H$,   $\phi_d$ obtained from ${\cal
A}_{\rm CP}(B_d \to J/\Psi K_s)$ and CKM-angle $\gamma$ (see 
\cite{FlMa2} for details). They
also depend on the two  U-spin
breaking parameters:  $\xi$ and $\Delta \theta$. We start the 
analysis in the
U-spin limit ($\xi=1,\Delta \theta=0$) and we  explore
in Sec.3.2 the sensitivity of the results to deviations from 
this
limit. An interesting remark is the symmetry\cite{FlMa2} that 
Eq.~(\ref{sp}) exhibits:
\begin{equation}\phantom{centeringspa}\label{sym}\phi_d
\rightarrow
180^\circ
-\phi_d \quad \gamma \rightarrow 180^\circ -
\gamma\end{equation}
The present world average
$\sin\phi_d=0.734\pm0.054$ gives rise to two possible solutions:
$\label{phid-det}
\phi_d=\left(47^{+5}_{-4}\right)^\circ \, \lor \,
\left(133^{+4}_{-5}\right)^\circ.
$
The first solution has  positive  $\cos\phi_d$  and the
second  negative  $\cos\phi_d$. Our approach allow us to
explore both.
These two solutions together with the symmetry of Eq.~(\ref{sym})
will have important consequences as we will see in a
moment.

\subsection{Determination of $\gamma$}

The experimental situation is still uncertain and the present
naive average is\cite{exp} (including PDG enlarged errors):
\begin{eqnarray} \phantom{center}
{\cal A}_{\rm CP}^{\rm dir}(B_d\to\pi^+\pi^-)&=&-0.51\pm0.19
\,\, (0.23) \nonumber
\label{Bpipi-CP-averages}\\
{\cal A}_{\rm CP}^{\rm mix}(B_d\to\pi^+\pi^-)&=&+0.49\pm0.27
\,\, (0.61) 
\label{exp}
\end{eqnarray}
Taking Eq.~(\ref{sp}) and varying ${\cal A}_{\rm CP}^{\rm 
mix}(B_d\to\pi^+\pi^-)$ in all the {\it positive} range, with 
$H=7.5$ 
for each 
solution
of $\phi_d$ we find\cite{FlMa2}:

\noindent \begin{itemize}
 \item
Fig. 1a corresponding to the solution $\phi_d=47^\circ$ is in 
good agreement with the usual CKM fits\cite{ckm} for $2 
\beta$, and 
it gives the
following prediction for $\gamma$:
\begin{equation}\label{scena}\phantom{centeringequation}35^\circ\lsim\gamma\lsim79^\circ 
\end{equation}
Moreover, it excludes gamma  values in the range of:
$88^\circ\lsim\gamma\lsim143^\circ $

\item Fig. 1b, on the contrary, corresponds to the
solution
$\phi_d=133^\circ$. This solution cannot be explained in the SM
context and requires the existence of New Physics. In this case
the prediction for $\gamma$ is:
\begin{equation} 
\phantom{centeringequation}\label{scenb}101^\circ\lsim\gamma\lsim145^\circ\end{equation}
Interestingly,  the corresponding excluded region in this case:
$37^\circ\lsim\gamma\lsim92^\circ$  overlaps precisely with the
preferred region of the CKM fits\cite{ckm} of the SM.

\end{itemize}

The symmetry between Fig. 1a and Fig. 1b is a consequence of 
Eq.~(\ref{sym}). Notice also that large values of   ${\cal 
A}_{\rm 
CP}^{\rm 
dir}$ can be
perfectly accommodated.

\begin{figure}[t]
\hbox to\hsize{\hss
\includegraphics[width=\hsize]{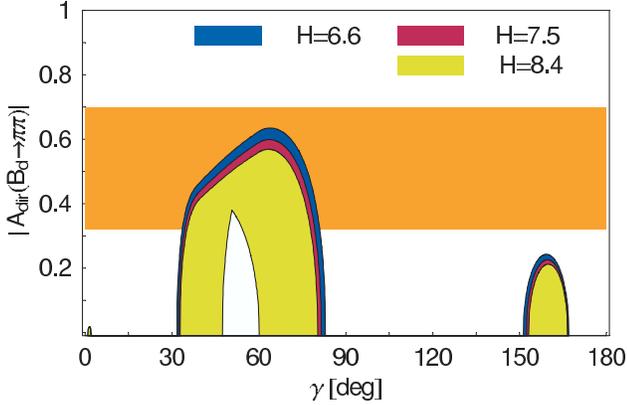}
\hss}
\label{fig7h}
\caption{Sensitivity to H}
\end{figure}
\begin{figure}[t]
\hbox to\hsize{\hss
\includegraphics[width=\hsize]{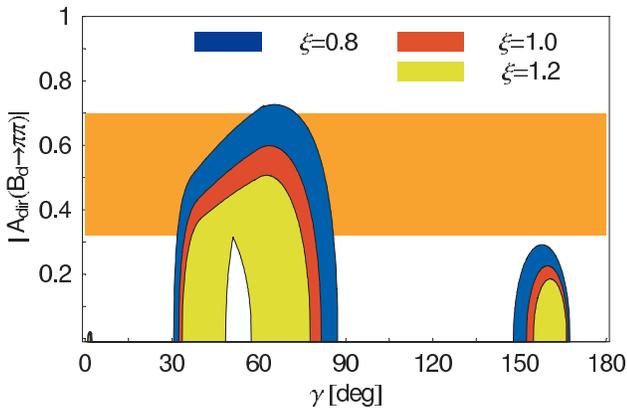}
\hss}
\label{fig7xi}
\caption{Sensitivity to $\xi$}
\end{figure}
%
%
%
\subsection{Sensitivity to H and $\xi$, $\Delta \theta$}

Here we examine  the sensitivity of CKM-angle  $\gamma$ to the 
variation of the different
hadronic parameters.

\begin{itemize}
\item H: Fig. 2 shows the change in the prediction for $\gamma$ when
H is varied between 6.6 to 8.4, for $\phi_d=47^\circ$. The region
shown corresponds to the restriction of ${\cal A}_{\rm CP}^{\rm mix}$ 
inside
the experimental range Eq.~(\ref{exp}). The error induced in the
determination of $\gamma$ is only of $\pm 2^\circ$. For the second solution
$\phi_d=133^\circ$ exactly the same conclusion can be drawn. One can enlarge the
range of H  as it was done in\cite{FlMa2} to take into account 
the uncertainty associated to the
spectator-quark hypothesis used to determine H, and the error
is still under control. Notice that with the future data on
$B_s \to KK$, this hypothesis {\it will not be 
needed}.

\item U-spin breaking parameter $\xi$. This is the most
important source of uncertainty. However, as can be seen in
Fig. 3, the error induced in the determination of $\gamma$ is $\pm
5^\circ$ even if we allow for a very large $\pm 20\%$ U-spin
breaking.

\item U-spin breaking parameter $\Delta \theta$. The effect on
the determination of $\gamma$
for values of $\Delta \theta$  up to
$40^\circ$ is completely  negligible.

\end{itemize}
Other studies on the use and evaluation of U-spin
can be found in\cite{flavour}.

\subsection{Determination of $\alpha$ and $\beta$ in the SM and
in presence of New Physics only in the mixing }

So far so good for $\gamma$, next question is how to
determine $\alpha$ and $\beta$.
Here we will also allow  for Generic New Physics affecting
the
$B^0_d$--$\overline{B^0_d}$ mixing, but not to the $\Delta 
(B,S)=1$
decay amplitudes.
In order to do so, we will use
three inputs\cite{FIM,GNW}:
\begin{itemize}
\item ${R_b}\equiv\left|\frac{V_{ud}V_{ub}^\ast}{
V_{cd}V_{cb}^\ast}\right|
$ obtained from exclusive/inclusive transitions mediated by  $b\to u
\ell\overline{\nu}_\ell$ and
$b\to c\ell\overline{\nu}_\ell$. Two important remarks are:
a) It is not expected that New Physics can affect significantly
this quantity, b) already from $R_b^{\rm max}=0.46$ we can
extract a maximum possible value for $\beta$: $|\beta|_{\rm
max}=27^\circ$.

\item $\gamma$ obtained as discussed in  Sec.3.1.

\item $\phi_d$ from  ${\cal A}_{\rm
CP}^{\rm mix}( B_d\to J/\psi K_{\rm S})$ is used as an input for
the CP asymmetries of $B_d \to \pi \pi$, but NOT to determine
$\beta$, since we assume that New Physics could be present.
For the same reason also $ \Delta M_d$ and
$\Delta M_s/\Delta M_d$ are not used as inputs.

\end{itemize}

Using these inputs we obtain two possible determinations for
$\alpha$
and $\beta$, corresponding to two different scenarios.

\begin{figure}[t]
\hbox to\hsize{\hss
\includegraphics[width=\hsize]{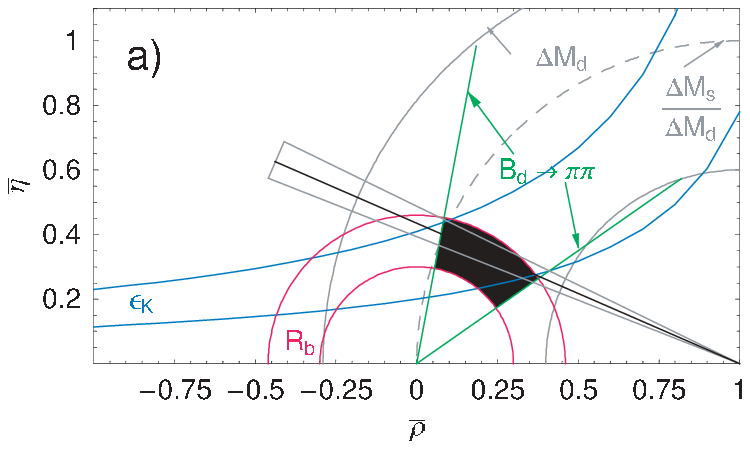}
\hss}
\includegraphics[width=\hsize]{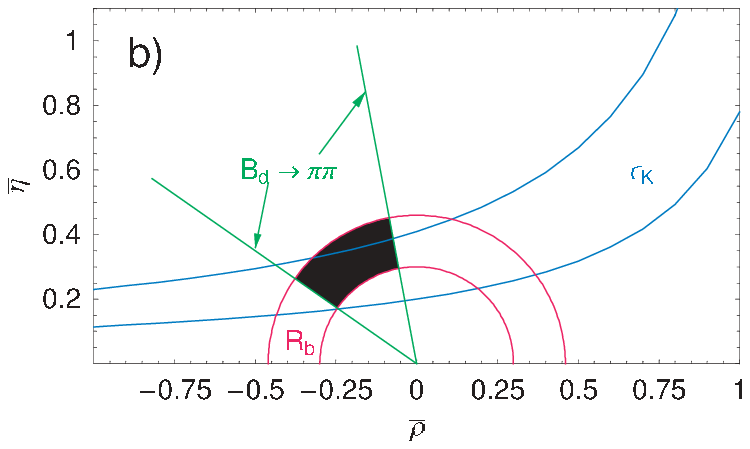}
\caption{a) SCENARIO A and b) SCENARIO B (see text).}
\label{fig:cathedral}
\end{figure}

SCENARIO A: Taking the determination of $\gamma$ from the CP asymmetries of $B_d \to
\pi \pi$ given by Eq.~(\ref{scena}), together with $R_b$ and
the first solution $\phi_d=47^\circ$
from
${\cal A}_{\rm CP}(B\to J/\Psi K_S)$ as an input for
the CP asymmetries of $B_d \to \pi
\pi$,
we
obtain the black region of Fig. 4a. This region, is in good
agreement with the usual SM
CKM fits\cite{ckm}. To illustrate this we shown in Fig. 4a also 
the
prediction from the SM interpretation of
different
observables: $\Delta M_d$, $\Delta M_s/\Delta M_d$, $\epsilon_K$ and
 $\phi_d^{SM}=2 \beta$.
The prediction for the CKM angles that we obtain using our method is:
$$74^\circ\leq\alpha\leq 132^\circ \quad
13^\circ\leq\beta\leq 27^\circ \quad
35^\circ\leq\gamma\leq79^\circ$$
and the error associated with $\xi\in[0.8,1.2]$ is $\Delta 
\alpha=\pm4^\circ$, $\Delta
\beta=\pm1^\circ$ and $\Delta \gamma=\pm5^\circ$.

SCENARIO B: The second solution: $\phi_d=133^\circ$ {\it cannot } be explained in the
SM context and requires New Physics contribution to the 
mixing\cite{GNW}. Several models, and
in particular supersymmetry\cite{bec}, in the framework of the
mass-insertion approximation\cite{mi}  can generate the extra
contribution to the mixing as it
was shown in\cite{FIM}. In this case, Fig. 4b, $\gamma$
lies in the second quadrant Eq.~(\ref{scenb}) and 
$\beta$ is indeed smaller than in the
previous scenario. The result is still consistent with the $\epsilon_K$ hyperbola.
$\Delta M_{d,s}$ are not shown here, since they would be affected by New Physics. The
black region obtained corresponds to the following prediction 
for the CKM angles:
$$24^\circ\leq\alpha\leq64^\circ \quad 8^\circ\leq\beta\leq22^\circ \quad
101^\circ\leq\gamma\leq145^\circ$$
with same errors associated to $\xi$ as in Scenario A.
It is interesting to notice that this second solution gives a better agreement with
data for certain very rare decays like $K^+\to\pi^+\nu{\bar
\nu}$\cite{FIM,rising,bb} than the SM
solution.

In conclusion the method described allow us to determine the CKM
angles using flavour symmetries with data from non-leptonic B
decays and $R_b$. The method is valid for the SM and in 
presence
of New Physics not affecting the amplitudes. Finally, the method
provides self-consistency checks to control the impact of hypothesis and
ways
{\it to eliminate}  some of them (spectator-quark hypothesis) when data
from hadronic machines will be available.

{\it Acknowledgements} JM acknowledges f.s. from
FPA2002-00748.
Very special thanks to Muntsa G. and J\'ulia M.

\end{document}